# A Novel Reward Shaping Function for Single-Player Mahjong

Kai Jun Chen, Lok Him Lai

*Department of Aeronautics and Astronautics, Stanford University, CA, USA*

Zi Iun Lai

*Department of Mechanical Engineering, Stanford University, CA, USA*

**Mahjong is a complex game with an intractably large state space with extremely sparse rewards, which poses challenges to develop an agent to play Mahjong. To overcome this, the ShangTing function was adopted as a reward shaping function. This was combined with a forward-search algorithm to create an agent capable of completing a winning hand in Single-player Mahjong (an average of 35 actions over 10,000 games). To increase performance, we propose a novel bonus reward shaping function, which assigns higher relative values to synergistic Mahjong hands. In a simulated 1-v-1 battle, usage of the new reward function outperformed the default ShangTing function, winning an average of $1.37 over 1000 games.**

## I. Introduction

Originating in 19th century China, mahjong is a turn-based 4 player game using ceramic tiles. Once derided as a pastime of the lazy and indulgent [1], mahjong has grown in popularity in recent years, due to cross cultural media platforms and the game's distinctive shuffling feature [2]. Scientific interest in mahjong first grew when reports suggested mahjong might delay cognitive decline in persons with dementia [3]–[5]. In the field of artificial intelligence (AI), mahjong attracts interest as a model problem due to its multiplayer and imperfect information features, and wide range of scoring variations [6]. It is interesting to note that at the time of this writing, the performance of top AI players has not surpassed human players in the game of mahjong. This project aims to develop a program to play a simplified game of mahjong.

## II. Game Mechanics

Mahjong rules vary considerably across geographic regions, and the rules are often very complex. In this project, we use a simplified version of mahjong, which consists of only the suited tiles and honors (wind and dragon) tiles, shown in Figure 1b. There are 4 identical copies of each tile, for a total of 136 tiles.

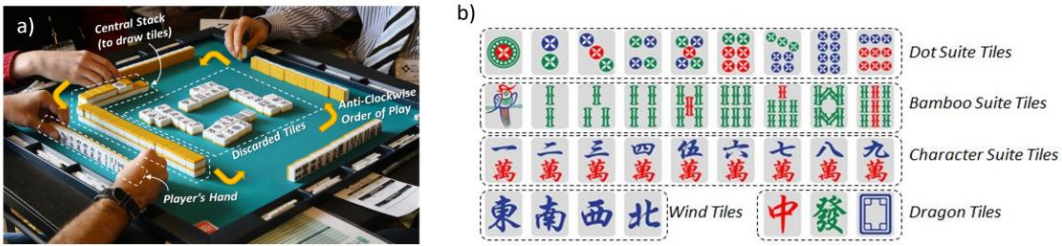

*Figure 1 – (a): Typical mahjong game setup, adapted from [7]. (b): Tiles in a simplified mahjong game.*

The layout of a conventional game is depicted in Figure 1a. Starting with 13 tiles each, players take turns to draw one tile (the drawn tile is random) and discard either that tile or any tile in hand. A player's tiles (including drawn tile) are not know to others, while the discarded tiles are visible to everyone. More details on Mahjong can be found here.

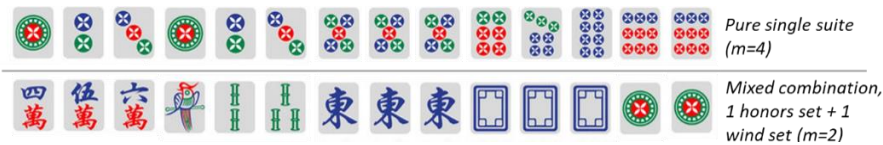

*Figure 2 – Examples of winning hand combinations*

The objective of the player is to win as much money as possible in the least number of turns. A player wins the game by completing a hand of 14 tiles consisting of a valid combination of sets and pairs. Some examples for winning hand combinations are shown in Figure 2. Different tile combinations provide different a multiplier, $m$. From the base payoff, $b$ (this amount is agreed to before the game, e.g., $2), the individual payoff is computed using $2^m b$. The 3 losing players each pay this amount, and the winning player receives a total payoff $3*(2^m b)$. Players thus have to balance between aiming for a common winning hand (thus winning first but receiving a smaller payout) against trying out for a rarer winning hand (with potentially greater rewards).

Mahjong's complexity lies in its large state space and possible changes in policies midway through the game [8]. At every stage, a player has 14 different actions corresponding to which tile to discard, and up to 34 unique state transitions, which are possible tiles to be drawn. Additionally, the tiles the player draws might change his/her policy on which winning hand to aim for, or when winning seems unlikely, which tile to avoid discarding so as to not benefit opponents. In this project, we target the single-player Mahjong problem, which functions as a building block.

## III. Literature Review

The state space of Mahjong is intractably large, and winning hands are a miniscule subset. Cheng et al. used combinatorics to determine the probability of a specific winning hand known as 'Nine Gates' [6]. This winning hand consists of all tiles of the same suite, drawing another any of the nine tiles from that suite forms a winning hand. They considered all 93600 possible combinations of drawing 13 tiles from a pile of 36 (suite tiles only) and reported that probability of obtaining 'Nine Gates' is 0.0113%. They further calculated the probabilities of obtaining 'X Gates' for X = 1 to 8. However, exhaustive calculation of probabilities of winning hands are usually not feasible. Online methods are therefore recommended over offline methods in Mahjong games. Several studies reported the use of deep learning approaches. Wang et al. fed game states and actions of top mahjong players into ResNet and Inception+ architectures, which won 28% of matches in four player games [9]. In 2020, Li et al. used supervised learning from logs of human players and reinforcement learning with policy gradient methods and encoded player's tiles into 2D array [10]. Their program achieved top 0.01% ranking on an online site. Koyamada et al. then implemented RL with C++ and to develop a faster API for mahjong AI [11]. Finally, tackling a 3-player variant of Mahjong, Zhao et al. used convolutional neural networks and RL to develop an AI with more aggressive playing strategies [12].

Mizukami et al [13] first abstracted Mahjong games and used supervised learning to determine a move for obtaining a winning hand, given a set of 14 tiles. They then incorporated other strategies (avoiding discarding winning tiles for opponents, rather than winning) into Monte Carlo Tree Search (MCTS) for multiplayer mahjong [14]. Zhang et al [17] proposed using a single-player Mahjong algorithm to handle the initial phase of the game, where the optimal action typically does not depend much on the actions of opponents. A switch to MCTS only happens in end-game stages, to predict the actions of opponents and estimate their possible hands. This strategy of using single-player Mahjong as a building block to full Mahjong was also adopted by [13].

## IV. Problem Statement

In this project, our team sought to tackle the single-player Mahjong problem, where a sole player discards and draws tiles until a winning hand is obtained. The objective is to develop a model that can discard appropriate tiles to attain a winning Mahjong hand in the least number of rounds while maximizing score. While we initially attempted the methods described in literature (neural network and MCTS), our limited computational resources precluded their effective use. We therefore focused on development of novel reward functions for a greedy strategy through incorporation of game heuristics.

## V. Methods

### A. Problem Formulation as MDP

The game state is initialized by randomly selecting 14 tiles to be in the player's hand. At every step, the player selects the index of tile to discard, and receives a random undrawn tile. The reward is a function of state only. The game continues until a winning hand is achieved or all tiles are drawn.

We convert the "single-player Mahjong" problem into a Markov Decision Process (MDP) problem (Figure 3):
- **State space (S):** One state is defined by a 2D array with 136 entries (equal to the number of tiles). The array consists of 34 columns (per tile), and 4 rows (per copy number). Each tile thus has a unique index. Array values are assigned '0', '1' or '2' depending on whether the tile is in the draw pile, player's hand, or discard pile respectively. The size of the state space is $|S| = \frac{136!}{14!(136-14)!} \times 2^{136-14}$.

- **Action space (A):** Actions 1 – 14 correspond to the tile to discard in that state, so it is dependent on the state. There are 14 unique actions. Note that an action includes both a discard and a random draw.
- **Transition model (T):** When a discard action is performed, the tile's value in the state array is changed from '1' to '2' (discarded). Then, a new tile is randomly drawn, and its value is changed from '0' to '1'. Since we assign a uniform probability across undrawn tiles, the transition model is known exactly.
- **Reward model (R):** Accepts a state array and returns a score based on Mahjong rules. Additionally, we assign a shaping function that provides small rewards for intermediate states that build up toward a winning hand. These tiny rewards will incentivize the policy towards winning hands but are small in magnitude compared to a winning score based on Mahjong rules.

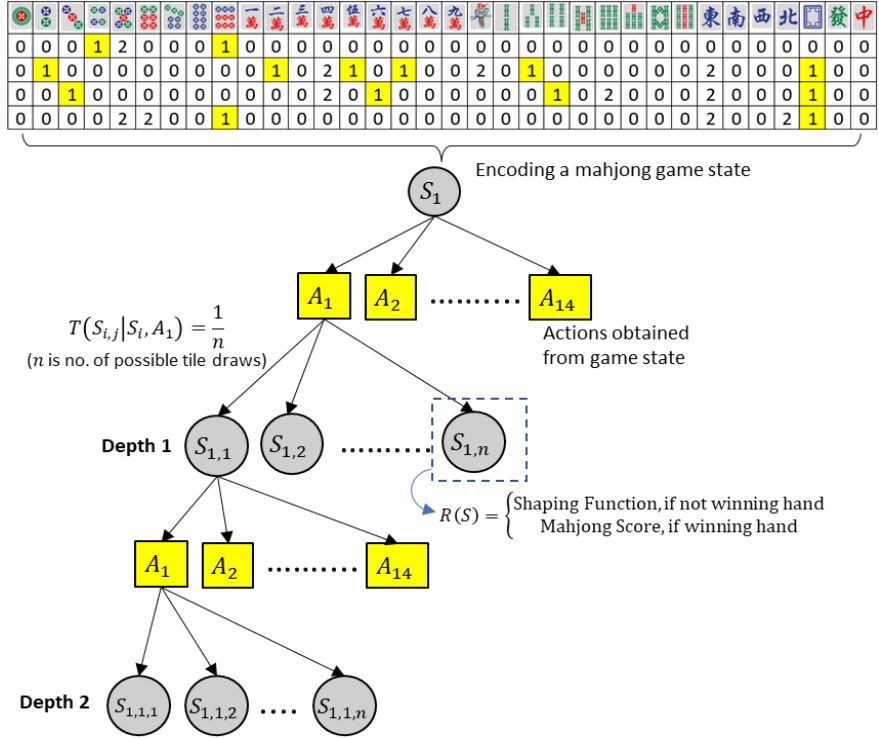

*Figure 3 – Converting the Mahjong problem to a Markov Decision Process*

### B. Scoring Functions

For a Mahjong game, the obvious reward function is to assign a score to states based on the official rules, if it is a winning hand (terminal state), and zero otherwise. We implemented this in the "*score_hand*" function that performs a series of logical checks based on Mahjong scoring rules.

Our initial attempts involved using a Monte Carlo Tree Search algorithm to solve the MDP, as has been examined in literature for many board games[15], [16]. Unfortunately, our search depth is very limited as computation grows exponentially with depth. For example, at the start of the game, a search up to depth 1 requires enumerating over 14 actions*122 states = 1708 leaf nodes, while a search up to depth 2 increases this to 2.9 million. In general, performing a full forward search up to depth n requires enumerating over $14^n \prod_1^n \frac{122!}{(122-n)!}$ leaf nodes. Since only the terminal state has a non-zero reward, the probability of randomly stumbling across a state that has a reward is extremely tiny. In all other situations, the Bellman backup equation would return zeros, which offers no value to our decision model.

To make the problem tractable, we propose reward shaping [18]. For Mahjong, we implemented the 'ShangTing' score [17] as the reward shaping function. The 'ShangTing' function encodes distance away from forming a winning hand using the following procedure:

1) Every hand is initialized with a score of -14.
2) Number of identical triplets are counted, and each triplet adds a score of 3. Triplets are then disregarded for further computation.
3) Number of consecutive tiles of the same suite are counted, and each series adds a score of 3. Series tiles are similarly disregarded.
4) Among the remaining tiles, the first pair further increments the score by 2. Subsequent pairs do not increase score.

A ShangTing distance of '0' denotes a winning hand, while a ShangTing distance of '-14' represents the hands that are the furthest distance away from a winning hand. One potential pitfall of this is that searching to further depths causes rewards to be repeatedly summed up. Intuitively, the completion of a winning hand should be independent of how many discards we had to make. To overcome this, we modify the ShangTing distance to be the difference between the current state and the next state. This would instead reflect the expected relative improvement of each potential action. When a wining hand is obtained, "*score_hand*" function is implemented to obtain the official Mahjong score.

### C. Solving the MDP

The MDP defined in parts A and B is solved using an iterative 1-depth forward search. We adopted this method for several reasons:
1) Mahjong's intractable state space requires an online planning method.
2) An exact update of action-state pair provides a more accurate estimate than randomly sampling over possible states. A large number of samples is needed to approach exact values, and an exact update allows the model to capture tiny differences in action-value functions between different actions.
3) The shaping function allows the model to evaluate intermediate states, such that even a 1-depth forward search is not being too greedy, since it also considers the expected potential improvement of next states.
4) Performing forward search to 2-depth and beyond requires exponentially more computation time.
5) Recalculating the transition function with each updated state accounts for the sequential nature of Mahjong by eliminating already discarded tiles from the pool of possible tiles to draw.

To implement this algorithm, we first initialize a game of Mahjong and find ourselves in a new, random game state ($s$). We compute the expected action-value functions ($Q$) for all possible actions ($a$) by summing over all potential next states using the transition model ($T$) and reward shaping function ($R_s$). Then, we formulate our policy ($\pi$) by selecting the action that maximizes the action-value function. We repeat this process until our reward shaping function identifies a winning hand, signifying a terminal state. Since a bonus reward function is implemented as mentioned previously, a weighting factor ($w$) is added to adjust the level of risk. The algorithm is implemented in Python, and the code attached in the appendix.

$$Q(s,a) = \sum_{s'} T(s'|s,a) R_s(s',s)$$

$$R_s(s',s) = \begin{cases} [\text{ShangTing} + \text{Bonus}](s',w) - [\text{ShangTing} + \text{Bonus}](s,w) & \text{(if not winning hand)} \\ \text{Official Mahjong Score} & \text{(if winning hand)} \end{cases}$$

$$\pi(s) = \underset{a}{\text{argmax}}\, Q(s,a)$$
$$s \leftarrow \text{Update state}(s, \pi(s))$$

### D. Novel Reward Shaping Function

To maximize earnings, we propose a new reward function that includes the ShangTing function and a so-called "*unscented_bonus*" component. The ShangTing component assigns scores to features which completes a basic winning hand (triplets, consecutives and pairs). With the "unscented bonus" portion, features such as having all tiles in the same suit, dominance of triplets over consecutive tiles, or having triplets of dragon tiles are incentivized. In mahjong, completion of winning hands with these features results in higher payoffs. A comparison of 500 games with ShangTing only and ShangTing with "unscented bonus" reward functions suggest the latter yields higher earnings with >99% confidence (appendix). Our "*unscented bonus*" function is computed by summing the following and multiplying by a scaling parameter or weight:
1) Assign a score of 3 to every triplet of wind or dragon tiles.
2) Assign a score of 1 to the remaining first instance of a wind or dragon tile pair.
3) Assign a score of 3*(max(number of tiles of a one suit)/(number of tiles of all suits))

We further compared the unscented function against 'black box' scoring functions. Using logs of 900 games, we trained a simple 3-layer neural network (NN) and extreme gradient boost regressor (XGBoost) to as reward functions. In addition to being significantly faster, incorporation of heuristics led to the novel reward function consistently outperforming NN and XGBoost models in 1-v-1 matches.

**E. 1-v-1 Simulations**

To further determine the performance of unscented bonus weights, modified 2-player matches were conducted. Two mahjong games with differently weighted reward functions were simulated in tandem. The match ends when one of the players achieve a winning hand; there was therefore a need to balance degree of exploration (in holding out for a larger payout) vs exploitation (in quickly forming a hand with smaller payout).

## VI. Results & Discussion

A snapshot of the solver in action is shown in Figure 4. The first row represents the player's hand, and the number below each tile is its $Q$-value. According to the model, the recommended tile to discard is "6-suo", because it has the highest $Q$-value. For the reader's viewing pleasure, a video of a full run of the Mahjong solver completing a winning hand can be found here.

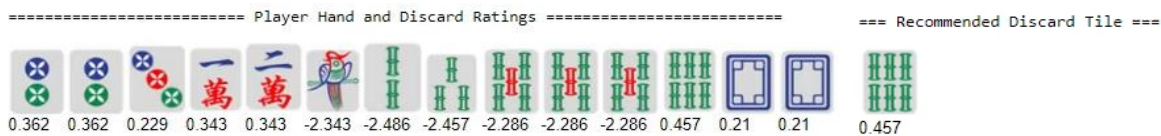

*Figure 4 – Snapshot of the solver after 13 actions in a particular Mahjong game*

Additionally, we ran the Mahjong solver for 10,000 games and successfully obtained a 100% completion rate. Detailed statistics are shown in Table 1 and Figure 5. On average, it took about 5 seconds to complete a winning hand, which our team believes is a reasonable run time. One interesting observation was that the number of discards to complete a hand had a very large variation, which implies that there is a great deal of luck in Mahjong. There was one game where only 3 discards were needed to complete a winning hand, while another game required 114! (The draw pile has 122 tiles).

|  | No. of discards to win | Avg. run time per full game (s) |
|---|---|---|
| **Mean** | 34.6 | 5.3 |
| **Min** | 3 | 0.31 |
| **Max** | 114 | 1052 |
| **Std. Dev.** | 18.2 | 50.1 |

| Mahjong Score | Count (% of games) |
|---|---|
| 1 | 86.70 |
| 2 | 12.06 |
| 3 | 0.251 |
| 4 | 0.782 |
| 5 | 0.093 |
| 6 | 0.056 |
| 7 | 0.065 |

*Table 1. Statistics over 10,000 games, using ShangTing reward function.*

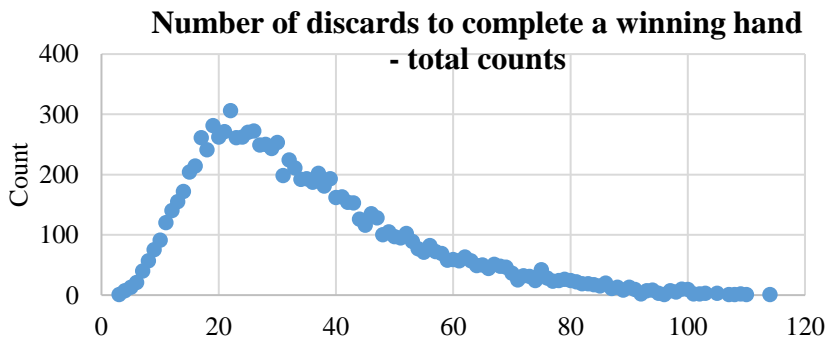

Figure 5. Number of Discards to Complete a Wining Hand

The weight of the unscented bonus function can be interpreted as a measure of how explorative an agent is. A weight of 0 denotes when the player uses the ShangTing function only in deciding which tile to discard and aims to obtain a basic winning hand. Increasing the unscented bonus weight causes players to value possible attainment of high-scoring hands more. Figure 6 presents earnings of player with unscented weight of 1.2 against weight of 0. The positive trend indicates that, over sufficient games, the use of unscented bonus function leads to net earnings.

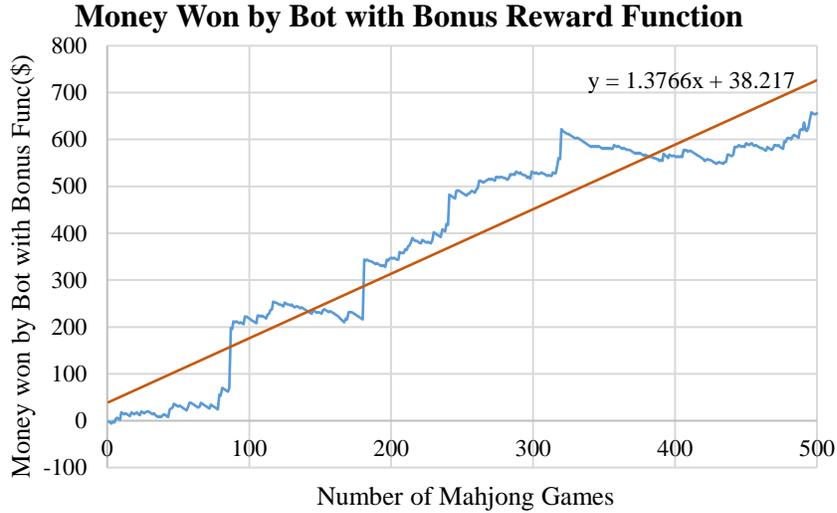

*Figure 6: Trend of winnings of unscented reward function with weight 1.2 against default ShangTing function (weight of 0), over 500 games.*

Reward functions of different weights were pitted against each other. In these runs, a positive number denotes a net increase in monetary earnings for player 2. Table 2 suggests the optimal weight is not unimodal and depends on the opponent's reward function. For instances, if the weight difference between player 1 and 2 is significant, the less explorative player tends to win more games. This is because the more cautions player can often form a winning hand earlier than the more explorative one, who takes bigger risks for higher payout. On the other hand, if their weight difference is closer, while the more explorative player tends to more games, the payouts from wins are often larger, leading to net increase in earnings.

|  |  | Player 2 weights | | | |
|---|---|---|---|---|---|
| Player 1 weights |  | 0.75 | 1.2 | 1.4 | 1.6 |
|  | 0.75 |  |  | +70 | -42 |
|  | 1 | -86 | +28 | +68 | -48 |
|  | 1.2 |  |  | -46 |  |
|  |  | Values denote winnings of player 2. | | | |

*Table 2: Outcomes for 1v1 games between players of different weights. Values denote earnings by player 2. Values for each combination are obtained after 100 simulated games.*

## VII. Conclusion

A one-player variation of Mahjong game was abstracted into an MDP problem. We propose a novel reward function to evaluate potential discard actions, using a combination of ShangTing distance and 'unscented bonus' functions. The agent's risk-taking behavior can be controlled through choice of 'unscented bonus' weight. Using a simple forward search strategy of depth 1, adoption of our reward function leads to long-term earnings against ShangTing-only agents. In 1-v-1 matches, the optimal 'unscented bonus' weight is likely multimodal, and depends on the opponent's playing style.

## VIII. Group Member Contributions

Kai Jun Chen: MDP formulation, novel reward shaping, report compiling, data collection, ideation.

Lok Him Lai: Shangting score function, report compiling, XGBoost exploration, visualizations & display functions, ideation.

Zi Iun Lai: Literature review, report compiling, data collection, NN exploration, confidence testing.

## Appendix

### A. Hypothesis testing

We'll use one tail test to determine if one scoring method is better than another. Let original hypothesis be that the improvement $\mu$ using the new method is less than or equal to 0. We believe the new method to be better.

$$h_0: \mu \leq 0$$
$$h_1: \mu > 0$$

We reference one-sided t value table for 1% significance level, with 500 samples. The t value corresponding to critical value is 2.334.

We then calculate $t_c$ value from our samples.

Comparing the match results between Unscented weight of 1.2 and default ShangTing, we calculate $t_c$ value with:

$$t_c = \frac{\bar{x} - \mu_0}{\sigma/\sqrt{n}} = 2.867$$

Where $\bar{x}$ is mean improvement, $\mu_0$ is improvement in original hypothesis, $\sigma$ is standard deviation of improvement and $n$ is the number of matches simulated. Since $t_c > t$, we reject the null hypothesis with 99% confidence. $t$ values can be found from tables such as in ref [19].

### B. Python Code

https://github.com/ckaijun0/mahjong

## Acknowledgments

We would all like to take this chance to express our profound gratitude to Professor Mykel J. Kochenderfer for teaching introducing us to the world of partially observable Markov Decision Processes.